\documentclass{aa}  

\usepackage{txfonts}
\titlerunning{A monolithic collapse origin for the thin/thick disc structure of ESO~243-49}

\begin{document}

   \title{A monolithic collapse origin for the thin/thick disc structure of the S0 galaxy \object{ESO~243-49}\thanks{Based on observations made at the European Southern Observatory using the Very Large Telescope under programme 60.A-9328(A).}}

   \author{S.~Comer\'on
          \inst{1}, H.~Salo\inst{1}, R.~F.~Peletier\inst{2},
          \and
          J.~Mentz\inst{2}
          }

   \institute{University of Oulu, Astronomy Research Unit, P.O.~Box 3000, FI-90014, Finland\\
             Kapteyn Astronomical Institute, University of Groningen, PO Box 800, NL-9700 AV Groningen, the Netherlands\\
             }
 
  \abstract{ESO~243-49 is a high-mass (circular velocity $v_{\rm c}\approx200\,{\rm km\,s^{-1}}$) edge-on S0 galaxy in the \object{Abell~2877} cluster at a distance of $\sim95$\,Mpc. To elucidate the origin of its thick disc, we use MUSE science verification data to study its kinematics and stellar populations. The thick disc emits $\sim80\%$ of the light at heights in excess of $3\farcs5$ (1.6\,kpc). The rotation velocities of its stars lag by $30-40\,{\rm km\,s^{-1}}$ compared to those in the thin disc, which is compatible with the asymmetric drift. The thick disc is found to be more metal-poor than the thin disc, but both discs have old ages. We suggest an internal origin for the thick disc stars in high-mass galaxies. We propose that the thick disc formed either {\it a)} first in a turbulent phase with a high star formation rate and that a thin disc formed shortly afterwards, or {\it b)} because of the dynamical heating of a thin pre-existing component. Either way, the star formation in ESO~243-49 was quenched just a few Gyrs after the galaxy was born and the formation of a thin and a thick disc must have occurred before the galaxy stopped forming stars. The formation of the discs was so fast that it could be described as a monolithic collapse where several generations of stars formed in a rapid succession.}


   \maketitle
%

\section{Introduction}

The discs of some galaxies have been divided into a thin and a thick component since \citet{BURS79} and \citet{TSI79}. Thick discs are better seen in close to edge-on galaxies, where the vertical structure of discs can be studied. In fact, thick discs have been found in most of the edge-on galaxies where they have been searched for \citep{YOA06, CO11A} with a few possible exceptions among low-mass galaxies \citep{CO11C, STREICH2016}. The Milky Way is also know to have a thick disc \citep{GIL83}.

In our work on photometric decompositions of edge-on galaxies we proposed that stars in thick discs have an internal origin, at least for galaxies with circular velocities $v_{\rm c}\gtrsim120\,{\rm km\,s^{-1}}$ \citep[from now on we will call those high-mass galaxies;][]{CO11C, CO12, CO14}. However, obtaining direct evidence of an internal origin requires thick disc kinematics. This reason is that thick discs formed via external accretion would result, at least in some galaxies, in a fraction of counter-rotating stars acquired in retrograde encounters.

Due to the technical difficulties arising when obtaining spectroscopy of faint features, only seven stellar thick disc rotation curves are known in high-mass galaxies, namely those in FGCE~1498 and FGCE~1371 \citep{YOA08}, that in ESO~533-4 \citep{CO15}, that in NGC~3115 \citep{GUE16}, and those in NGC~4111, NGC~4710, and NGC~5422 \citep{KAS16}. The thick discs of all those galaxies have prograde rotation curves that lag at most a few tens of kilometers per second with respect to the thin discs. The Milky Way thick disc is also prograde \citep{CHI00}. On the other hand, the few observed rotation curves in low-mass galaxies have a larger variety than those in high-mass galaxies. For example, some of them show little net rotation \citep{YOA05, YOA08}.

If thick discs were accreted, their stars would have been obtained in encounters with small impact angles with respect to the disc. This is because dynamical friction would drag such low inclination satellites towards the plane of the galaxy. Satellites with a large impact angle would have their stars ending in the stellar halo \citep{READ08}.

An important question to study is to find out what the origin of thick discs is, i.e., internal or external. If the accretion of stars through several minor mergers were creating thick discs, one would expect to always find retrograde stars in them because roughly similar amounts of prograde and retrograde mergers would occur in a given galaxy \citep{READ08}. However, if thick discs were formed in a small number of encounters, there would be no guarantee for a given galaxy to have a retrograde thick disc. In the limit in which thick discs were created in a single merger, there would be similar chances to have a retrograde or a prograde encounter. Thus, based on the kinematics of the before-mentioned eight thick discs a major merger origin cannot be ruled out as the main thick disc formation mechanism in high-mass galaxies. To completely discard it, it is important to increase the number of known thick disc rotation curves.

The MUSE instrument at the VLT \citep{BA10} is an integral field unit with a large field of view (FOV) of $1^{\prime}\times1^{\prime}$, which makes it ideal for the study of nearby galaxies. ESO~243-49 was observed during the science verification phase. The whole galaxy fits within the MUSE FOV. ESO~243-49 was classified as an S0 galaxy by \citet{DRE80}. The MUSE data show that the galaxy has a conspicuous round bulge and no evidence of recent intense star formation, which indeed makes it an S0 galaxy. It is located at a distance of $\sim95\,{\rm Mpc}$ if we consider a heliocentric radial velocity of $V_{\rm Helio}=6667\,{\rm km\,s^{-1}}$ \citep{JO09} and a Hubble-Lema\^itre constant $H_0=70\,{\rm km\,s^{-1}\,Mpc^{-1}}$. ESO~243-49 is a member of the Abell~2877 cluster and hosts the ultra-luminous X-ray source HLX-1, which is the most likely candidate for an intermediate mass black hole \citep{FAR09}.

\section{Preparation of the science data-cube}

The MUSE data of ESO~243-49 are made of 16 exposures (8100\,s total exposure). For each exposure a data-cube was created using the MUSE reduction pipeline within the Reflex environment \citep{FREUD13}. The target did not occupy the whole FOV, so we set the pipeline to create a sky model using spaxels outside the galaxy (15\% of all spaxels). Then, the 16 data-cubes were manually aligned. We combined them using the \texttt{muse\_exp\_combine} recipe. The sky residuals of the combined data-cube were removed with ZAP \citep[][]{SO16}. The wavelength range of the final data-cube is $4750-9350\AA$.

\section{Thin/thick disc decomposition}

\label{fit}

\begin{figure}
\begin{center}
  \includegraphics[width=0.48\textwidth]{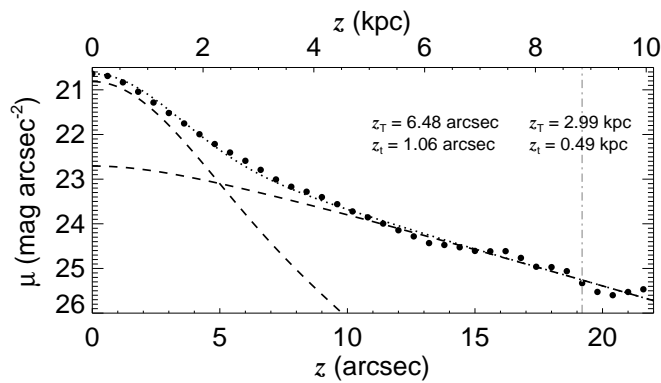}
  \end{center}
  \caption{\label{profile} 3.6$\mu$m vertical surface brightness profile for regions between 10 and 30$\arcsec$ from the galaxy centre (large dots). The dotted line denotes the fit to the profile and the dashed lines denote the thin and thick disc contributions. $z_{\rm T}$ and $z_{\rm t}$ indicate the thick and thin disc scale-heights. The grey dot-dashed line indicates the limit of the height range over which the fit was made.} 
\end{figure}

We obtained a $3.6\mu$m image of ESO~243-49 from the Spitzer Heritage Archive. We then created a surface brightness profile of the galaxy perpendicular to the mid-plane. We only considered the regions between 10 and 30$\arcsec$ from the centre (Fig.~\ref{profile}) because we ignored the bulge region (that where velocity dispersion $\sigma$ is about or larger than $100\,{\rm km\,s^{-1}}$ in Sect.~\ref{velocity}). We fitted the profile with the thin/thick disc decomposition code used in \citet{CO11C, CO12, CO15}. The code assumes two coupled vertically isothermal stellar discs. To do the fit, we used the 2D point spread function (PSF) from \citet{SA15}.

Making a thin/thick disc fit requires to know the ratio between the mass-to-light ratios of the thick and the thin discs  \citep[$(\Upsilon_{\rm T}/\Upsilon_{\rm t})_{3.6\mu{\rm m}}$;][]{CO11C}. To estimate that ratio it is necessary to know the ages and the metallicities of the stellar populations. In Sect.~\ref{pop} we find that both discs are old but have different metallicities ($\left[Z/H\right]=0.1$ for the thin disc and $\left[Z/H\right]=-0.15$ for the thick disc). We used the spectral energy distribution libraries (SEDs) from \citet{BRU03} to obtain the mass-to-light ratios of such populations at $3.6\mu$m. Linear interpolation was used to find SEDs intermediate between two metallicity templates. We considered an age of 13\,Gyr for both discs. We found that the mass-to-light ratio of the thick disc is 1.04 times that of the thin disc ($(\Upsilon_{\rm T}/\Upsilon_{\rm t})_{3.6\mu{\rm m}}=1.04$). Using that information in our thin/thick disc fit we find that the thick disc is 0.58 times as massive as the thin disc ($\Sigma_{\rm T}/\Sigma_{\rm t}=0.58$).

The $\Sigma_{\rm T}/\Sigma_{\rm t}$ estimate has several uncertainty sources. One is of course the determination of $(\Upsilon_{\rm T}/\Upsilon_{\rm t})_{3.6\mu{\rm m}}$. $\Upsilon_{3.6\mu{\rm m}}$ changes little with age or metallicity for such old populations. Changes of 0.1\,dex in metallicity or 1\,Gyr in age would produce changes smaller than 10\% in $(\Upsilon_{\rm T}/\Upsilon_{\rm t})_{3.6\mu{\rm m}}$. Since $\Sigma_{\rm T}/\Sigma_{\rm t}$  roughly scales with $\Upsilon_{\rm T}/\Upsilon_{\rm t}$ \citep{CO12}, the errors in the disc mass determination associated with the determination of $\Upsilon_{3.6\mu{\rm m}}$ would be of that order also. This is smaller than the $\lesssim20\%$ estimate of the error caused by approximations such as considering exactly the same radial scale-length for the thin and thick discs or ignoring the vertical gravity field due to dark matter halo in the fit \citep{CO12}.

Later on in this paper, we use the concept of thick disc dominated regions (Sect.~\ref{pop}). We define those regions as those at heights above $3\farcs5$. Above this height, about 82\% of the light in the MUSE data-cube comes from the thick disc. This number has been calculated by convolving the $3.6\mu$m fitted disc components with a Gaussian with a $0\farcs8$ full width at half maximum (FWHM) obtained from fitting the radial luminosity profiles of stars in the MUSE FOV. The FWHM varies between $0\farcs7$ and $0\farcs9$ within the instrument wavelength range. We also assumed that on average, at MUSE wavelengths $(\Upsilon_{\rm T}/\Upsilon_{\rm t})_{\rm MUSE}=0.81$. Again, these mass-to-light ratios are based on the SEDS from \citet{BRU03}. 

We compared the fit results to those of other early type galaxies (stages $-3 \leq T \leq 1$) with similar masses (circular velocity $v_{\rm c}=200\pm40\,{\rm km\,s^{-1}}$; see Sect.~\ref{velocity} for the $v_{\rm c}$ determination) in our previous studies. Four of such galaxies are found in \citet{CO12}. Those galaxies have $\Sigma_{\rm T}/\Sigma_{\rm t}$ values between 0.3 and 0.7, which is similar to what we find here.

\section{The velocity field of ESO~243-49}

\label{velocity}

\begin{figure}
\begin{center}
  \includegraphics[width=0.48\textwidth]{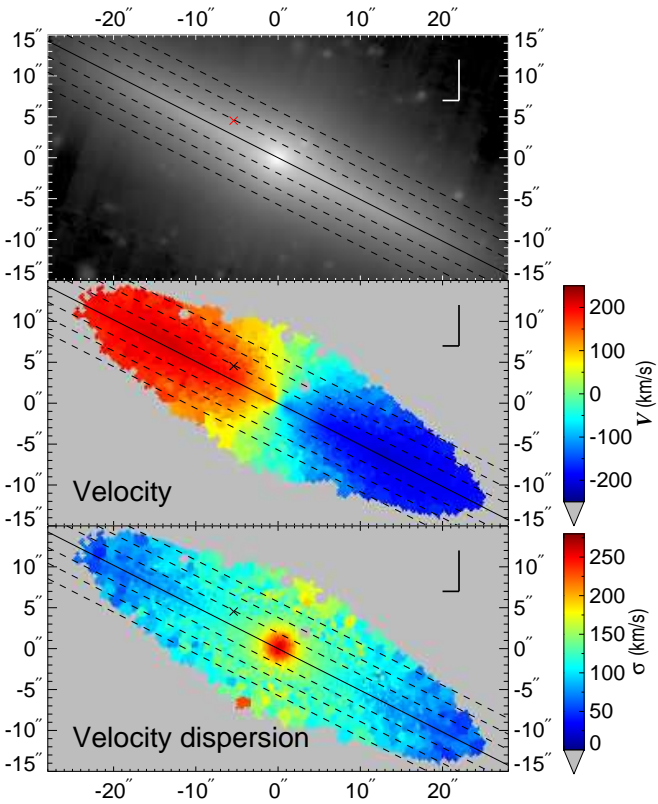}
  \end{center}
  \caption{\label{mosaic} {\it Top}: image of ESO~243-49 made by collapsing the MUSE data-cube along its spectral direction. {\it Middle}: radial velocity map. {\it Bottom}: velocity dispersion map. The cross some $7\arcsec$ to the NE of the centre of the galaxy indicates the location of the X-ray source HLX-1 based on the information in \citet{MAP13}. The continuous line indicates the mid-plane and the dashed lines indicate heights of 0.8, 1.6, and 2.4\,kpc above and below the mid-plane.} 
\end{figure}

\begin{figure}
\begin{center}
  \includegraphics[width=0.48\textwidth]{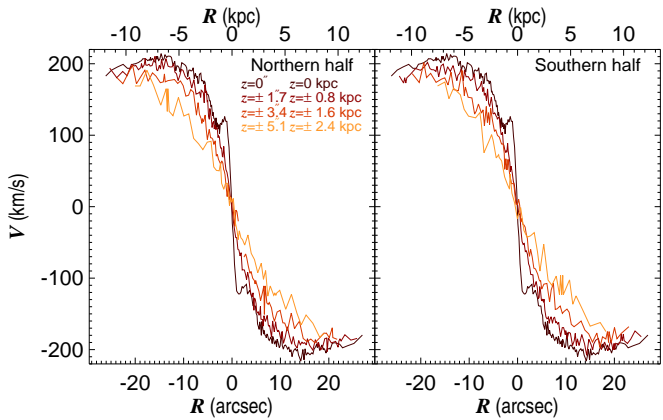}
  \end{center}
  \caption{\label{profiles} Velocity curves of ESO~243-49 for different heights above ({\it left}) and below the mid-plane ({\it right}).} 
\end{figure}

\begin{figure}
\begin{center}
  \includegraphics[width=0.48\textwidth]{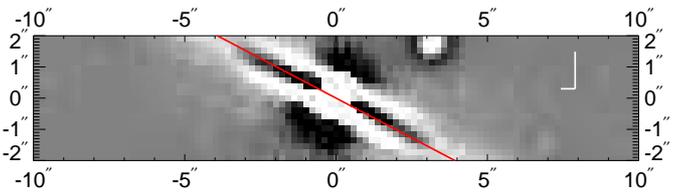}
  \end{center}
  \caption{\label{unsharp} Unsharp mask of the central regions of ESO~243-49. The red line indicates the mid-plane of the galaxy.} 
\end{figure}

We produced a Voronoi tesselation of the combined data-cube with the \citet{CAP03} software. We required a mean signal to noise ratio $S/N=25$ per pixel in the range between $5490\AA$ and $5510\AA$. We only considered spaxels for which $S/N>2$ in that wavelength range. The number of bins is 3705.

To obtain the line of sight velocity distribution (LOSVD) of each of the bins, we fitted the $4750-8800\AA$ range with the MIUSCAT stellar population synthesis models \citep{VAZ12} using the penalized pixel-fitting (pPXF) software by \citet{CAP04}. An eight order additive Legendre polynomial was used to model the continuum. We masked the regions with sky emission lines with peak emissions larger than $2\times10^{-18}\,{\rm W\,m^{-2}\,nm^{-1}}$ in \citet{HA03}. The mask width was $1000\,{\rm km\,s^{-1}}$. We masked the location of possible emission lines using the same mask width. We also masked the ${\rm O}_2-{\rm A}$ telluric band with a $4400\,{\rm km\,s^{-1}}$ wide mask. We only fitted two momenta (velocity, $V$, and velocity dispersion, $\sigma$) because higher momenta would require higher $S/N$ which would preclude us from getting a high spacial resolution in low surface brightness regions.

An image, the velocity map, and the velocity dispersion map of ESO~243-49 are shown in Fig.~\ref{mosaic}. The canonical butterfly diagram of a rotating disc appears in the velocity map. The innermost $2^{\prime\prime}$ region has a small butterfly pattern that corresponds in size with the dust lane at the galaxy centre (Fig.~\ref{unsharp}). The butterfly thus indicates a nuclear disc. That very same region has velocity dispersions in excess of $\sigma=200\,{\rm km\,s^{-1}}$ which points at the presence of a classical bulge as also indicated by the roundness of the inner isophotes. We therefore suggest that ESO~243-49 has a composite bulge, like those studied in \citet{FAL06} and \citet{ER15}. The classical bulge might also cause the relatively high velocity dispersions of $\sim150\,{\rm km\,s^{-1}}$ $3\,{\rm kpc}$ above and below the centre of the galaxy.

The rotation curve of ESO~243-49 at different heights above and below the mid-plane is shown in Fig.~\ref{profiles}. The mid-plane rotation curve shows two peaks at projected radii $R=\pm2^{\prime\prime}$ that correspond to the nuclear disc. The rotation curve becomes flat at about $R=6^{\prime\prime}$ (2.8\,kpc) with a circular velocity $v_{\rm c}\approx200\,{\rm km\,s^{-1}}$. ESO~243-49 is therefore a high-mass galaxy whose mass is slightly below that of the Milky Way. The maximum velocity goes down as the height increases. At a height of $5\farcs1$ (2.4\,kpc) - well into the thick disc dominated region - the stars lag by $\sim30-40\,{\rm km\,s^{-1}}$ with respect those in the mid-plane.

Is a $\sim30-40\,{\rm km\,s^{-1}}$ lag compatible with pure asymmetric drift or could it also include some effect due to a not perfectly edge-on orientation and/or to a fraction of retrograde stars? The dust lane in the innermost regions of ESO~243-49 runs exactly through the mid-plane of the galaxy (Fig.~\ref{unsharp}), which implies that it is not very far from edge-on and the lag is not due to projection effects. Also the lack of vertical asymmetries in the velocity map supports a close to edge-on orientation \citep[see][for a galaxy where a $2-3\degr$ deviation from an edge-on orientation a significant impact in the velocity map]{CO15}. To figure out whether the lag is compatible with the absence of retrograde stars, we compared it with that found in a galaxy with a similar mass, i.e. the Milky Way. Several studies point to a lag in the range of $40-60\,{\rm km\,s^{-1}}$ for the thick disc of the Milky Way \citep{OJ94, SOU03, PAS12, KOR15}. More detailed studies that describe the asymmetric drift as a function of the height above the mid-plane find similar values for a height of $z=2.4\,{\rm kpc}$ \citep{BOND10, MO12}. Thus, the lag of the thick disc of ESO~243-49 does not require large amounts of retrograde stars to be explained.

HLX-1 has been suggested to be at the centre of the nucleus of an accreted galaxy \citep{MAP13}. Our kinematic maps show no sign of a disturbed rotation curve or increased velocity dispersion at the position of HLX-1 (though a small increase in the velocity dispersion could be hidden by the fact that HLX-1 is found at a position where the bulge still influences the velocity dispersion map). This means that either the hypothetical merger was quite minor or that it was coplanar with ESO~243-49 and prograde and hence contributed to the thick disc component.

\section{Stellar populations}

\label{pop}

\begin{figure}
\begin{center}
  \includegraphics[width=0.48\textwidth]{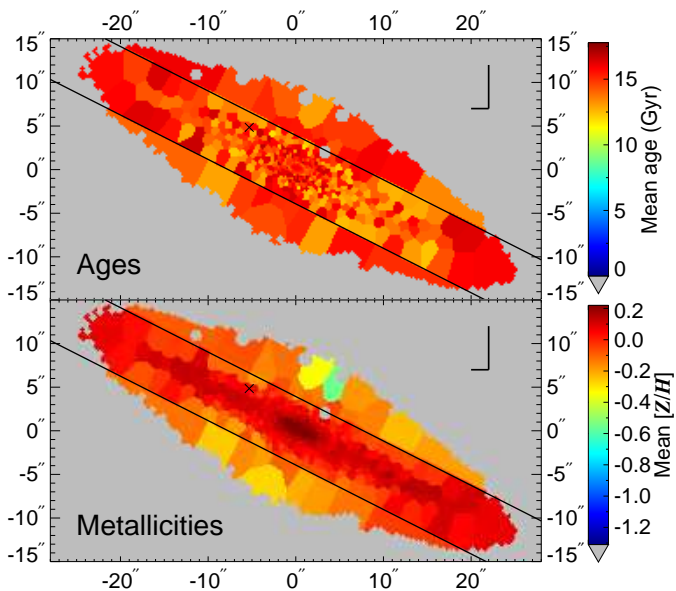}
  \end{center}
  \caption{\label{spmaps} {\it Top:} mass-weighted fitted mean age of the stellar populations. {\it Bottom:} mass-weighted fitted mean metallicity of the stellar populations. The solid lines indicate the height above which 82\% of the light is emitted by the thick disc according to the fit in Sect.~\ref{fit}. The cross indicates the location of HLX-1.} 
\end{figure}

\begin{figure}
\begin{center}
  \includegraphics[width=0.48\textwidth]{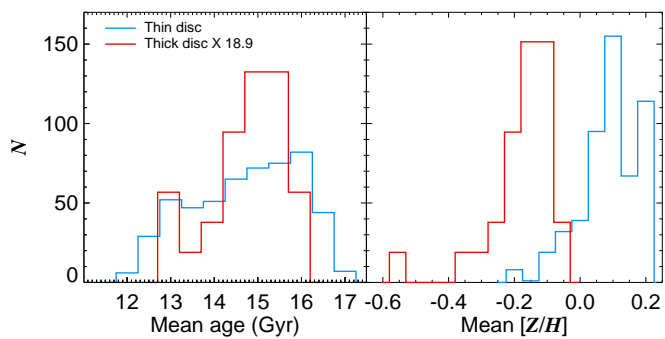}
  \end{center}
  \caption{\label{methist} {\it Left:} Distribution of the mean mass-weighted ages of the thin and thick disc bins (blue and red lines, respectively). {\it Right:} Distribution of the mean mass-weighted metallicities. The thick disc histograms have been multiplied to have the same surface as the thin disc ones.} 
\end{figure}

To study the stellar populations we created spacial bins with $S/N=80$. We run the Voronoi binning code three times. First we created a series of ``thin disc dominated'' bins in the section with heights smaller than 1.6\,kpc ($3\farcs5$; above that height 82\% of the light comes from the thick disc according to Sect.~\ref{fit}). Then we created ``thick disc dominated'' bins in the regions below (second time) and above (third time) the ``thin disc dominated'' area. The total number of bins was 558 (530 for the thin disc and 28 for the thick disc). We then fitted the LOSVD of each of the bins. This time we fitted four momenta ($V$, $\sigma$, and the Hermite momenta $h_3$ and $h_4$). For each of the bins, a subset of the MIUSCAT library (we excluded the templates with metallicities $\left[Z/H\right]<-1$) was widened by convolving it with the LOSVD. The widened templates were used to fit the spectra for each of the $S/N=80$ spacial bins with \texttt{STECKMAP} \citep{OC06B, OC06A}. The fit output was a star formation history (SFH) as well as the mean metallicity for each of the age bins of the SFH. The fit was done over the $4750-8800\AA$ wavelength range. The continuum was modelled with a spline function with 40 equi-spaced nodes.

Both the thin and the thick discs have old stellar populations with mean ages in excess of 12\,Gyr (Fig.~\ref{spmaps}). The two components have a similar mean age distribution (Fig.~\ref{methist}). No young stellar populations are found. Only six bins have more than 15\% of their light emitted by stars younger than 5\,Gyrs. On the other hand there is a significant vertical metallicity gradient with large metallicities in the mid-plane and lower metallicities in the thick disc. On average, thin disc bins have a slightly super-solar metallicity (peaking at $\left[Z/H\right]=0.1$), whereas the thick disc has a slightly sub-solar metallicity (peaking at $\left[Z/H\right]=-0.15$; Fig~\ref{methist}). The metal-rich band in the thin disc extends to heights about half of $3\farcs5$ so the PSF contamination of high-metallicity stars in the thick disc bins is likely to be minimal.

The trends found here are stable with respect to changes in the fitted wavelength range.

One might wonder whether our results are affected by the feared age-metallicity degeneracy. Using \texttt{STECKMAP}, \citet{SAN11} demonstrated that full spectral fitting techniques greatly reduce the degeneracy compared to what is obtained from the study of spectral indices. According to the plots in their paper they found that for a 10\,Gyr population with a solar metallicity the uncertainties are smaller than 1\,Gyr in age and about 0.2\,dex in metallicity. While this is indicative of the uncertainties due to the age-metallicity degeneracy in our fits, one has to take into account that our fitted wavelength range is not the same as in their study so the beforementioned values might not strictly apply (they used the $4150-6100\,\AA$ range whereas we use that between $4750$ and $8800\AA$).

An independent way to check the age uncertainties is to look at the scatter in the fitted age. This can be done because the age map does not show any clear age gradient, so it is reasonable to assume that all the spacial bins have a similar mean age. The standard deviation of the age distribution in Fig.~\ref{methist} is 1.2\,Gyrs.

The uncertainties described here apply to relative age and metallicity differences. Indeed, absolute age and metallicity determinations are affected by other uncertainties such those introduced by the choice of a given SED library. This probably explains that we find the discs to be slightly older than the age of the universe in standard cosmology (Fig.~\ref{methist}).

\section{Discussion and conclusion}

We find that ESO~243-49 has a thick disc with a lag compatible with that coming from asymmetric drift. It also has a composite bulge made of a classical bulge with a high velocity dispersion and a nuclear disc. The thick disc is not found to be older than the thin disc, but it is significantly less metallic.

Even though the ages of the thin and the thick discs are similar, the higher metallicity of the thin disc indicates a longer evolution that has led to an increased enrichment. The reason why this larger evolution period does not show in the age maps might be that the spectra of old stellar populations evolve little with time. Thus small age differences could be hidden by the uncertainties involved in the fitting process. It seems that the star formation in the thin disc of ESO~243-49 was quenched early on, probably because its presence inside the Abell~2877 cluster. ESO~243-49 might be a prototypical case of a galaxy with a monolithic collapse, where several generations of stars form very quickly one after another.

With this study, we now have the stellar kinematics of nine thick discs in high-mass galaxies. In all cases, the lag of the thick disc is of a few tens of ${\rm km\,s^{-1}}$. Based on our experience while studying ESO~533-4 \citep{CO15}, we estimate that a fraction of $10-20\%$ of counter-rotating stars in the thick disc would cause its rotation curve to lag too much compared to what would be expected from asymmetric drift. Therefore, our results suggest an internal origin for the majority of thick disc stars. The thick disc could have formed quick and in situ in a turbulent thick star-forming disc \citep{EL06, BOUR09, CO14} or in a slightly slower pace due to internal heating \citep{VILL85} or heating caused by interactions \citep{QUINN93}. Radial migration of stars could also contribute to the bimodal disc structure. This is because once the guiding centre of a star is modified, the amplitude of its vertical oscillations can change. This last possibility is the focus of a fierce debate \citep[see, e.g.][for different views on this]{SCHO09A, SCHO09B, LOEB11, MIN12, ROS13, VE14, GRAND16}.

The thin/thick disc differentiation should have happened in the first few Gyrs of ESO~243-49, when the galaxy was still forming stars, because posterior dynamical heating would tend to erase the metallicity gradients.

\begin{acknowledgements}
SC and HS acknowledge support from the Academy of Finland. We thank Lodovico Coccato for his help at refining the data reduction. We thank Jes\'us Falc\'on-Barroso and Inma Mart\'inez-Valpuesta for useful discussions. This research has made use of the NASA/ IPAC Infrared Science Archive, which is operated by the Jet Propulsion Laboratory, California Institute of Technology, under contract with the National Aeronautics and Space Administration. This research is partially based on data from the MILES project.

\end{acknowledgements}

\bibliographystyle{aa}
\bibliography{ESO243-49}

\end{document}